# Elucidating the topological nature of molecular beam epitaxy grown high Curie temperature $(Mn_{1+y}Sb_{2-y}Te_4)_x(Sb_{2-y}Mn_yTe_3)_{1-x}$ using angle resolved photoelectron spectroscopy

Candice R. Forrester[1,2], Jisoo Moon[1], Sina Mohammadi[3,4], Asish Kundu[5], Amy Lopez Navarro[1], Kaushini Wickramasinghe[6], Abhay Pasupathy[5,7], Anil K. Rajapitamahuni[5], Maya N. Nair[5,8], Elio Vescovo[5], Lia Krusin-Elbaum[4,9], Maria C. Tamargo[1,2,4,9, a]

*[1]Department of Chemistry, The City College of New York, NY, NY 10031, USA*

*[2]Ph.D Program in Chemistry, CUNY Graduate Center, NY, NY 10016, USA*

*[3]Department of Physics, Queens College of CUNY, NY, NY 10031, USA*

*[4]Ph.D Program in Physics, CUNY Graduate Center, NY, NY 10016, USA*

*[5]Brookhaven National Laboratory, Upton, NY 11973, USA*

*[6] Department of Physics, University of Peradeniya, Peradeniya, 20400, Sri Lanka*

*[7]Department of Physics, Columbia University, New York, NY 10027, USA*

*[8]CUNY Advanced Research Science Center, New York, NY 10031, USA*

*[9]Ph.D Program in Physics, CUNY Graduate Center, NY, NY 10016, USA*

a) **Electronic mail: mtamargo@ccny.cuny.edu**

## ABSTRACT

Angle-resolved photoemission spectroscopy (ARPES) is used to elucidate the band structure of $(Mn_{1+y}Sb_{2-y}Te_4)_x(Sb_{2-y}Mn_yTe_3)_{1-x}$, a new magnetic quantum material grown using molecular beam epitaxy that has a Curie temperature ($T_C$) above 100K, much higher than those reported for other materials of the $Mn_xSb_{2-x}Te_4$ class. Conflicting and sparse theory and experiments, as well as the complex disordered structure of the materials make it difficult to establish their topological nature. Here we explore this question by performing ARPES studies for two samples with x is 0.65 and 0.92, respectively. Images of the band structures of these two samples were acquired using photon energy dependent studies. The bulk and surface states of this complex material system, and the effects of composition on the band structure were investigated. Evidence of surface state features as well as bulk related ones were obtained, supporting their non-trivial nature. No signature of Fermi-arcs consistent with Weyl semimetals were seen. A shift in the Fermi level of the two samples due to different carrier densities allowed observation of the onset of a Dirac point in the surface states. Temperature dependent ARPES suggests possible band splitting as the materials undergo magnetic transitions, helping to understand the magnetic interactions. Our results point to their likely magnetic topological insulating nature.

## 1 Introduction

Topological materials are receiving much attention due to their unique exotic physical properties and their robustness to external perturbations and defects. When these characteristics are intertwined with magnetism, the interplay between these properties becomes an exciting frontier in which to observe and manipulate novel physical phenomena, including the quantum anomalous Hall Effect or chiral superconductivity[1] as well as the axion insulating or Weyl Semi-Metallic states.[2] Novel quantum devices for spintronics[3–5] and fault-tolerant quantum computing are anticipated due to coexistence of topology and magnetism.[6] The most studied of these materials are $MnBi_2Te_4$ and $MnSb_2Te_4$, the Mn containing derivatives of the 3D topological insulators (TIs) $Bi_2Te_3$ and $Sb_2Te_3$.[7–16] Their unique properties emerge from the introduction of magnetic ions, which breaks time reversal symmetry and modifies the band structure of the TI.[17] While $MnBi_2Te_4$ has been extensively studied with the consensus that it behaves as a magnetic topological insulator (MTI), its familial counterpart, $MnSb_2Te_4$, has garnered conflicting evidence and lack of consensus regarding its topological nature.[16,18–22] Early reports asserted that stoichiometric $MnSb_2Te_4$ is topologically trivial, where gapless inverted bands are observed, [15,22,23] or that it behaves as a Weyl semi-metal (WSM) with or without compressive strain, where inverted bands are also observed with discrete gapless points (Weyl nodes) forming a Fermi arc on the surface.[18,24,25] Many of the previously published works probed bulk stoichiometric $MnSb_2Te_4$ and extracted the crystal parameters in DFT calculations to predict the electronic band structure. However, Wimmer[16] and others,[23] asserted that $MnSb_2Te_4$ thin films were in fact magnetic topological insulators with a Curie temperature ($T_C$) of 50K due to excess Mn in the $MnSb_2Te_4$ lattice.

Incorporation of magnetic ions, such as Mn, into a 3D TI host can occur either as a structural element, yielding a new septuple layer (SL) crystal structure such as $MnSb_2Te_4$, or by substituting for the constituent atoms, forming quintuple layer (QL) alloys of the type $(Mn,Sb)_2Te_3$.[26] Both types of substitutions can affect the electronic and magnetic properties of the final structure. A large interest has risen around understanding the magnetic behavior of these materials. In stoichiometric $MnSb_2Te_4$, a single van der Waals SL is expected to be ferromagnetic, but stacked multiple layers align antiferromagnetically. It is in these bulk grown, near-stoichiometric materials that some groups have reported that the Weyl semi-metallic behavior emerges.[18,19,24,27] However, magnetic topologically insulating behavior in thin film $MnSb_2Te_4$ can be observed with excess Mn, which increases the amount of Mn/Sb site mixing;[20] In both bulk and thin film cases, ferromagnetic alignment was observed due to the innate Mn/Sb site mixing in $MnSb_2Te_4$.

Recently our group showed that we can achieve ferromagnetic materials with high Curie temperatures ($T_C$) for structures of $(MnSb_2Te_4)_x(Sb_2Te_3)_{1-x}$ by adjusting the composition x during molecular beam epitaxy (MBE) growth.[28,29] For x = 0.7 – 0.85, values of $T_C$ above 100K were achieved, far above those reported for similar materials.[28–30] Moreover, for samples within this regime, we proposed a new material: $(Mn_{1+y}Sb_{2-y}Te_4)_x(Sb_{2-y}Mn_yTe_3)_{1-x}$, in which significant amounts of excess Mn incorporate into the crystal, forming $Sb_{2-y}Mn_yTe_3$ QL alloys.[28] However, the topological nature of these structures has not been established. For example, density functional theory (DFT) and experiments reported by others have yielded conflicting results suggesting in some cases $MnSb_2Te_4$ is a Weyl semimetal (WSM), [21,24] topologically trivial,[31] or a TI.[13,16,20] Also, the high p-type carrier density in these samples makes it difficult to ascertain their band structure. Understanding the nature of these high $T_C$ materials[28,30] is essential for their potential application in practical spintronic quantum devices.

In this work, we investigated the effect of varying the ARPES measurement parameters such as photon energy, and the effect of tuning the Fermi level by modifying the composition (x) of the samples to elucidate the band structure of high $T_C$ $(Mn_{1+y}Sb_{2-y}Te_4)_x(Sb_{2-y}Mn_yTe_3)_{1-x}$. We further sought to study the band structure of these novel quantum materials as a function of temperature to observe changes in the band structure as the temperature varies across the $T_C$ values. Our results demonstrate that ARPES analysis can be used to obtain information about the topological nature and the effect of magnetization on the band structure of these complex and disordered materials. The experimental results support the non-trivial nature of the materials.

## 2 Experimental Details

Different compositions of $(Mn_{1+y}Sb_{2-y}Te_4)_x(Sb_{2-y}Mn_yTe_3)_{1-x}$, thin films were synthesized using a molecular beam epitaxy (MBE) Riber 2300P chamber system under ultra-high vacuum conditions (3-5 x $10^{-10}$ Torr). Each composition was achieved by adjusting the Mn beam equivalent pressure (BEP) ratio [Mn BEP ratio = Mn BEP/(Sb BEP + Mn BEP)] and the growth temperature; more details of the growth procedure can be found in refs. 28-30. The samples were grown on c-plane (0001) sapphire substrates and monitored in-situ using reflection high energy electron diffraction (RHEED) to ensure the quality of the film. Before growth, the sapphire substrate surfaces are prepared and cleaned via a thermal annealing at 650°C for an hour.

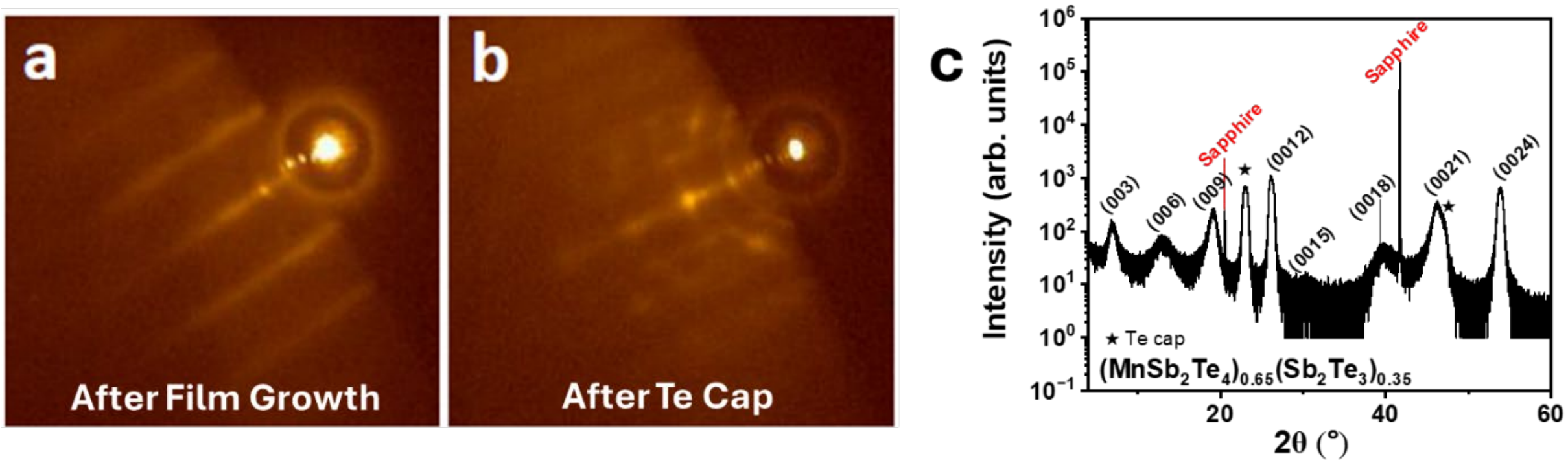

**Figure 1: Surface and structural characterization of $(Mn_{1+y}Sb_{2-y}Te_4)_{0.65}(Sb_{2-y}Mn_yTe_3)_{0.35}$ with Te cap**. a) Streaky RHEED image of $(Mn_{1+y}Sb_{2-y}Te_4)_{0.65}(Sb_{2-y}Mn_yTe_3)_{0.35}$ after 2-hour growth. b) Spotty RHEED image of Te capped layer after 10 minutes. c) HR-XRD spectrum of $(Mn_{1+y}Sb_{2-y}Te_4)_{0.65}(Sb_{2-y}Mn_yTe_3)_{0.35}$ with the Te cap (before desorption).

As the substrate is being treated, the BEP for 6 N antimony (Sb), 6N tellurium (Te) and 5N8 manganese (Mn) were measured and adjusted for a 1:20 Sb:Te ratio and 0.09 – 0.1 Mn BEP ratio. After the growth was completed, the surfaces of the films were prepared for transport in air between the MBE and ARPES chambers by depositing an amorphous Te layer at 15°C for 10 minutes within the MBE chamber. As seen in Figure 1a, right after growth, the RHEED patterns show streaky (1x1) surface reconstruction indicating good crystalline quality of the films. After the Te cap layer was deposited to protect the surface (before ARPES band structure analysis), the RHEED becomes noticeably spotty, and polycrystalline rings appear (Figure 1b). The X-ray 2θ-ω scans of the samples after Te deposition show clear presence of Te on the surface (starred peak, Figure 1c).

## 3 Results and Discussion

The band structure of a sample of $(Mn_{1+y}Sb_{2-y}Te_4)_x(Sb_{2-y}Mn_yTe_3)_{1-x}$ with $x \geq 0.65$ was studied to investigate the topological nature of these high $T_C$ structures. First, the composition was determined using XRD analysis. The quality and composition of the film was assessed by HR-XRD spectra. The 2θ-ω scan shown in Figure 1c is indicative of the quality of the sample. Using Vegard's Law to describe the shift in the (0021) reflection peak relative to pure $Sb_2Te_3$,[30] the structure measured in Figure 1c is expected to have a value of x = 0.65 (or 65%SL), which is described by the formula $(Mn_{1+y}Sb_{2-y}Te_4)_{0.65}(Sb_{2-y}Mn_yTe_3)_{0.35}$. Previous work has shown that the excess Mn in the crystal promotes higher $T_C$ values, particularly in those samples with x = 0.65-0.85.[28,29] The very high Mn content in the structures gave evidence of the formation of $(Sb_{2-y}Mn_yTe_3)$ alloys with y values as high as 0.6.[28]

### 3.1 Band structure of high $T_C$ in $(Mn_{1+y}Sb_{2-y}Te_4)_{0.65}(Sb_{2-y}Mn_yTe_3)_{0.35}$

Once the samples are transferred into the ARPES preparation chamber, the thin 30nm amorphous Te cap layer was desorbed by annealing the sample at 250°C for about 20 minutes. The cap removal was verified by conducting XPS measurements. The XPS data is seen in Figure 2a. No peak from the Te cap (elemental Te) is observed, expected to be at around -110 eV. As indicated in the expanded view of Figure 2a, the remaining Te peaks are related to Sb-Te bonds. Once the cap was removed, Fermi surface mapping of the $(Mn_{1+y}Sb_{2-y}Te_4)_{0.65}(Sb_{2-y}Mn_yTe_3)_{0.35}$ structure was probed at 18K, at a photon energy of 77eV. A six-fold petal contoured image around the Γ point was acquired, seen in Figure 2b. This feature is typical of a hexagonal crystal structure. Fermi arcs typical of a Weyl semi-metal, which were predicted [18,15,24,32] and observed[18,19,27] by others, were not observed in our highly disordered samples. Moreover, a few groups have studied how WSM phase or features may be affected in highly disordered systems. Their DFT calculations suggested that in strongly disordered systems, the Fermi arcs collapse and dissipate,[33] leaving a "metallic bath" where no distinguishing features are observed; it appears as a "blur".[33,34] This is not the case for the MST sampled studied in this paper where clear, six-fold petal-like Fermi surface is observed.[33,34]

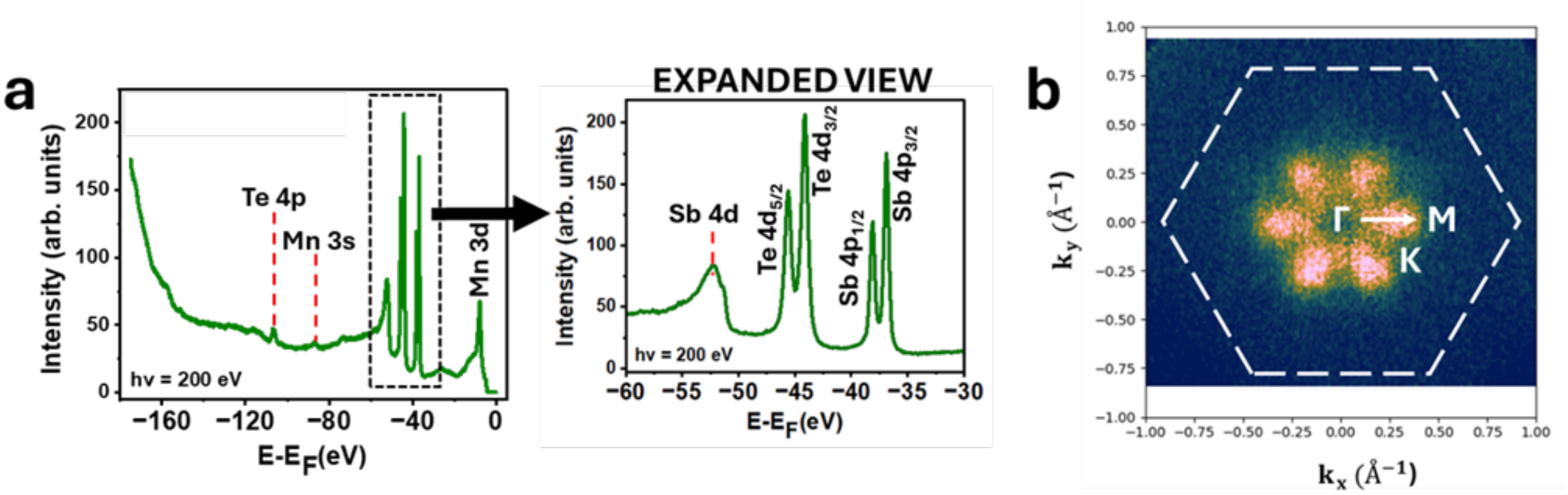

**Figure 2: a) XPS of $(Mn_{1+y}Sb_{2-y}Te_4)_{0.65}(Sb_{2-y}Mn_yTe_3)_{0.35}$ after the removal of Te cap layer.** The spectrum shows little to no evidence of elemental Te left, providing a good surface for ARPES analysis. *right*) Expanded view of the dashed rectangle, between -60 to -30 eV, identifies Te peaks associated with Sb-Te bonds. b) Fermi mapping of $(Mn_{1+y}Sb_{2-y}Te_4)_{0.65}(Sb_{2-y}Mn_yTe_3)_{0.35}$ surface after Te cap is removed revealing a six-fold symmetrical Fermi surface. The white arrow shows direction M-Γ-M cut used for band structure analysis.

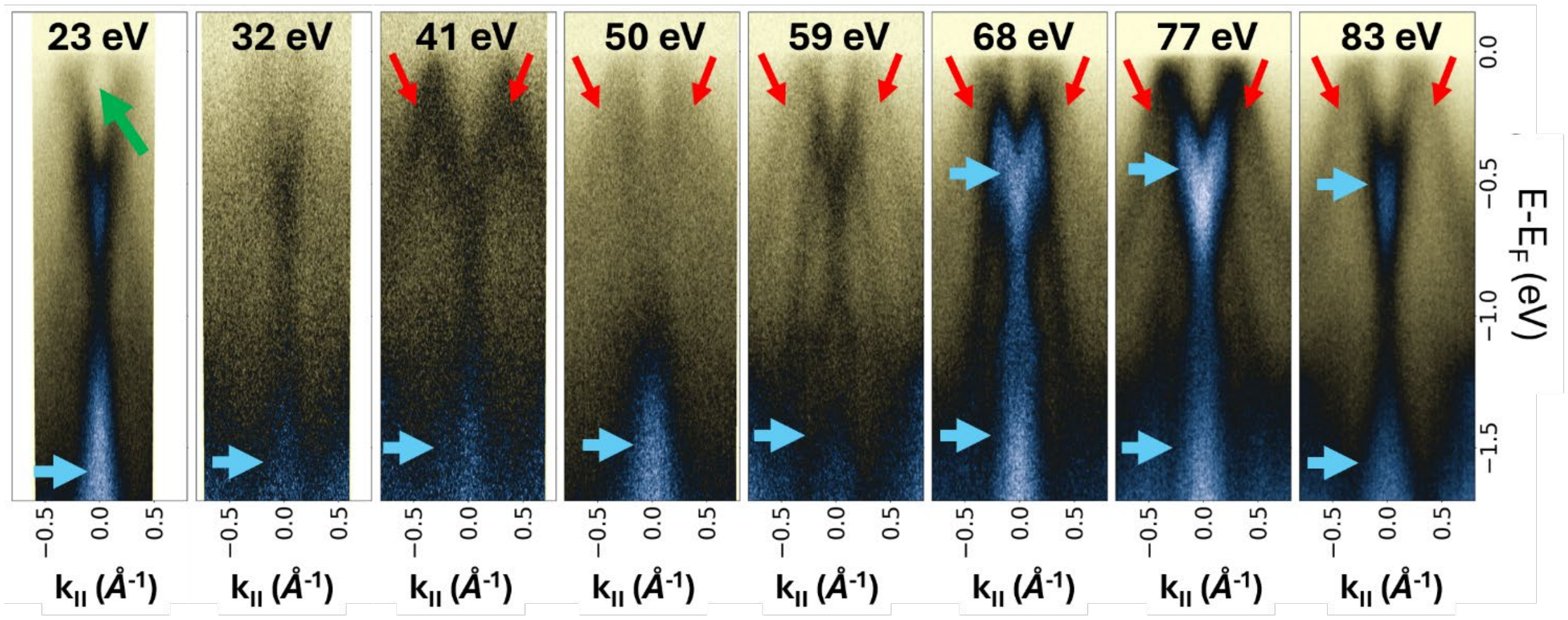

**Figure 3: Photon energy dependent study used to determine ideal photon energy to capture surface and bulk properties of $(Mn_{1+y}Sb_{2-y}Te_4)_{0.65}(Sb_{2-y}Mn_yTe_3)_{0.35}$.** The red arrows were used to identify topologically protected surface bands. Blue arrows draw attention to dispersion in bulk valence bands as a function of photon energy. The green arrow highlights possible hole pocket in band structure. (Following the notation in Ref. 13 and 16).

From the Fermi surface characterization, the path for the photon energy dependent study was selected as M-Γ-M. Along this path, several images of the band structure were collected between 23 eV and 83 eV, as seen in Figure 3. With a broad photon energy study, different depths of the crystal can be probed based on the universal mean free path: longer mean free paths (higher photon energy) are more sensitive to bulk features and shorter mean free paths (lower photon energy) are sensitive to surface features. Photon energy dependent studies of previously published work[16] suggest that $MnSb_2Te_4$ with excess Mn exhibits a band structure consistent with that of a magnetic topological insulator, which compare well with the results of our high $T_C$ $(Mn_{1+y}Sb_{2-y}Te_4)_{0.65}(Sb_{2-y}Mn_yTe_3)_{0.35}$ structure shown in Figure 3. Red and blue arrows are used to highlight different parts of the band structure in Figure 3, as previously done in Refs 13 and 16. The red arrows distinguish the unchanging, topologically protected surface bands, despite changes in the photon energy, and the blue arrows indicate the bulk bands that exhibit dispersion as a function of photon energy. This is further seen in Figure S1 of the Supplementary Material. The lack of dispersion in the surface bands is evidence of their 2D character and is consistent with the topological nature of these high $T_C$ structures. A possible hole pocket in the uppermost band structure, analogous to the one reported in reference 16 is also seen, identified by the green arrow.

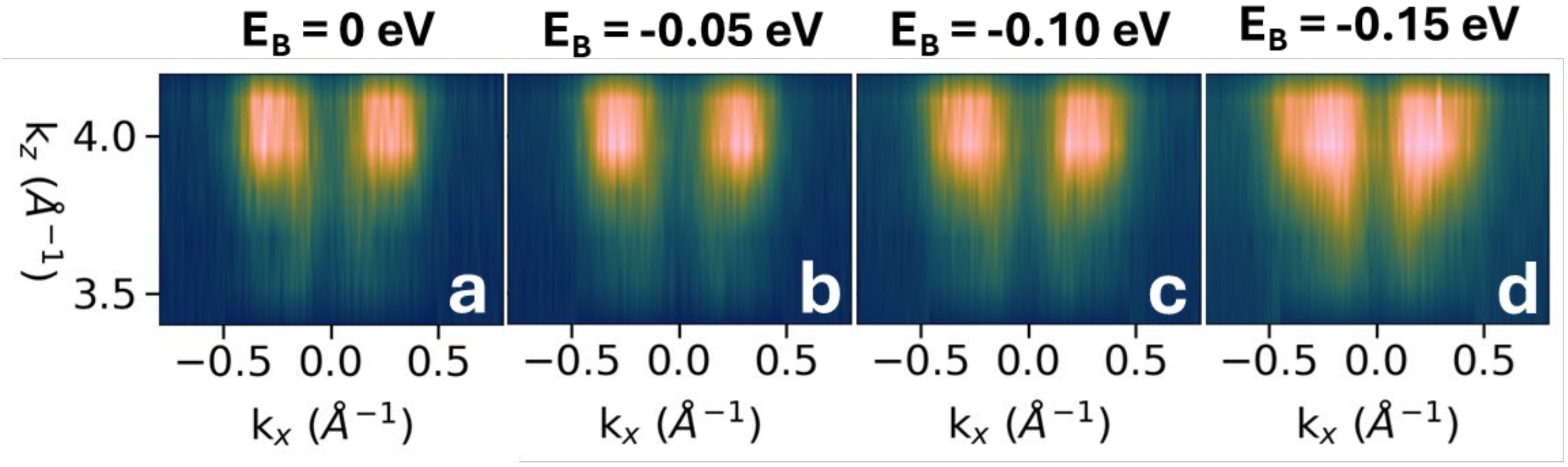

**Figure 4: Detailed analysis of hole pocket seen in Figure 3.** a-d) $k_z$ vs $k_x$ analysis of the hole pocket seen in Figure 3 at different binding energies. Little to no dispersion is observed, indicating that topological surface states are being probed.

The band structure was further analyzed to confirm the possible 2D character of the features. Figure 4 shows several images of the $k_x$ vs $k_z$ plots where the hole pocket is found, which were acquired at different binding energies (EB). The lack of dispersion (i.e., changes with binding energy) is also a feature of the topological quality of the surface states, which are not perturbed by external influences. No significant photon energy dependence ($k_z$) was observed in these plots, suggesting that we are probing the topological surface states.

### 3.2 Band structure of $(Mn_{1+y}Sb_{2-y}Te_4)_{0.92}(Sb_{2-y}Mn_yTe_3)_{0.08}$

A similar ARPES study was conducted on a higher composition sample $(Mn_{1+y}Sb_{2-y}Te_4)_{0.92}(Sb_{2-y}Mn_yTe_3)_{0.08}$. The composition was calculated using the position of the (0021) peak from the XRD scan, shown in Figure 5. XPS again provided evidence of the Te cap removal before proceeding with the ARPES study. Similar to the previous sample, Figure 5c shows Fermi surface mapping at Γ which presents a six-fold symmetry contour, more closely aligned to topological insulating or Dirac semi-metal behavior than with a Fermi arc typical of a WSM. The M-Γ-M path highlighted in white is the cut used to acquire images for the photon energy dependent study. However, the Fermi surface for this sample has a diffused and distorted contour, unlike the more regular six-fold shape seen in Figure 2b. This may be due to incomplete desorption of the cap. As seen in Figure 5b, the Te 4p peak is stronger than that observed in the $(Mn_{1+y}Sb_{2-y}Te_4)_{0.65}(Sb_{2-y}Mn_yTe_3)_{0.35}$ sample. Based on the study of Figure 3, in which the best observation of bulk and surface features occurred at 77eV, the band structure for this sample, $(Mn_{1+y}Sb_{2-}$ ${}_yTe_4)_{0.92}(Sb_{2-y}Mn_yTe_3)_{0.08}$, was measured between 69eV and 83eV. We observed the most features at 83eV, including the electron pocket highlighted by the green arrow.

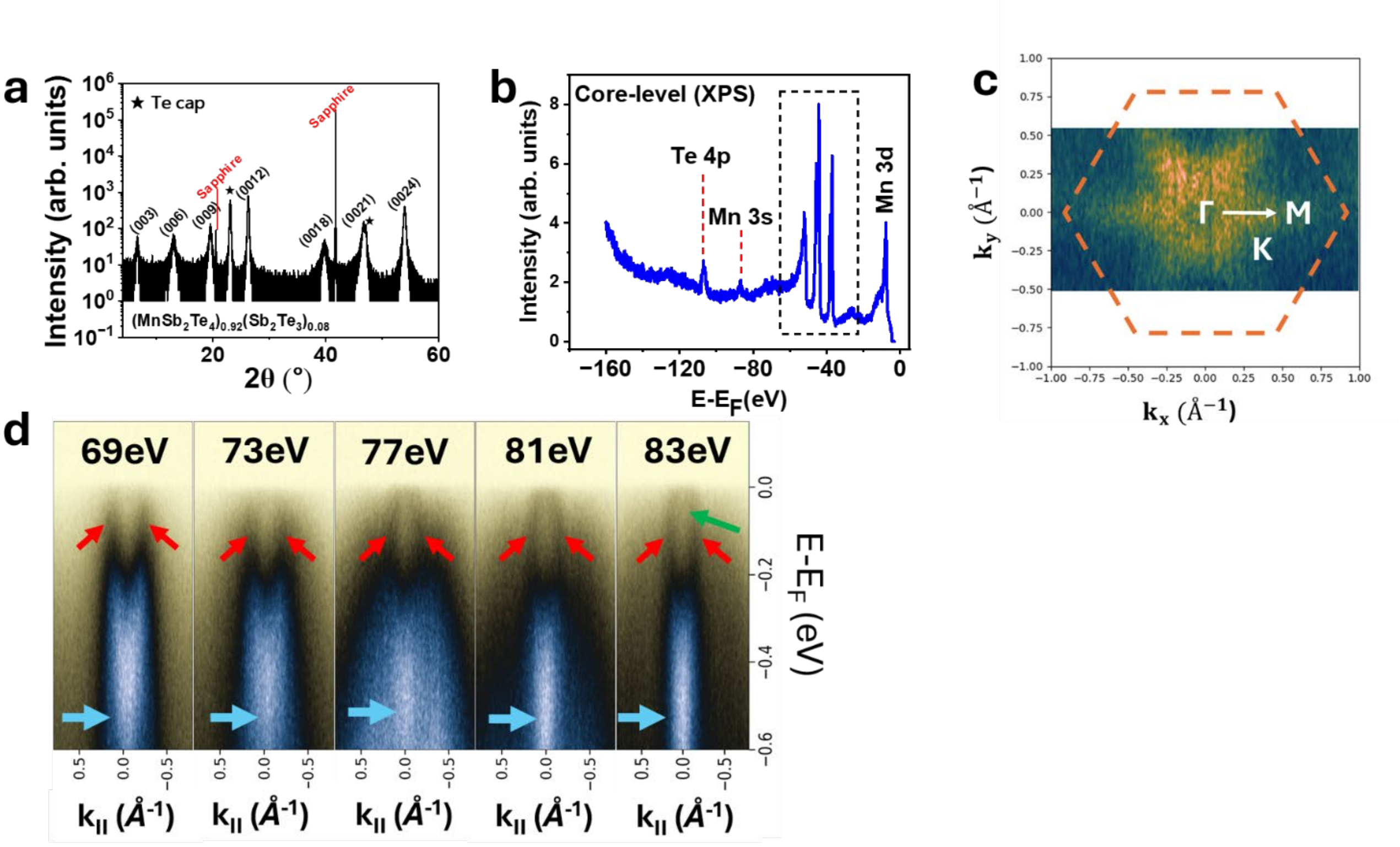

**Figure 5: Characterization of $(Mn_{1+y}Sb_{2-y}Te_4)_{0.92}(Sb_{2-y}Mn_yTe_3)_{0.08}$.** a) HR-XRD spectra of $(Mn_{1+y}Sb_{2-y}Te_4)_{0.92}(Sb_{2-y}Mn_yTe_3)_{0.08}$. Sharp peaks suggest highly crystalline material with a Te cap indicated by the star. b) XPS of $(Mn_{1+y}Sb_{2-y}Te_4)_{0.65}(Sb_{2-y}Mn_yTe_3)_{0.35}$ after the removal of Te cap layer. The spectrum shows little to no evidence of elemental Te left, providing a good surface for ARPES analysis c) Six-fold symmetry observed after removal of Te cap. d) photon energy dependent study of $(Mn_{1+y}Sb_{2-y}Te_4)_{0.92}(Sb_{2-y}Mn_yTe_3)_{0.08}$. Red and blue arrows highlight the surface and bulk bands, respectively. The green arrow highlights the hole pocket.

### 3.3 Tuning the Fermi Level

The $(Mn_{1+y}Sb_{2-y}Te_4)_x(Sb_{2-y}Mn_yTe_3)_{1-x}$ layers are typically p-type materials, in which the Fermi level is found within the valence band, making it difficult to probe the surface states in the gap. Several ways have been shown to control and manipulate the carrier density including varying the sample composition.[29,35–37] The carrier density of our structures as a function of $(Mn_{1+y}Sb_{2-y}Te_4)_x(Sb_{2-y}Mn_yTe_3)_{1-x}$ composition x is shown in Figure 6, where the blue points represent samples previously grown by us (see refs. 30 and 31) across the full compositional range of x and the red points represent the two new samples of this study: $(Mn_{1+y}Sb_{2-y}Te_4)_{0.65}(Sb_{2-y}Mn_yTe_3)_{0.35}$ and $(Mn_{1+y}Sb_{2-y}Te_4)_{0.92}(Sb_{2-y}Mn_yTe_3)_{0.08}$. The new samples show a similar behavior to what was previously observed, in which materials with higher values of x had a lower carrier density. In this study, $(Mn_{1+y}Sb_{2-y}Te_4)_{0.65}(Sb_{2-y}Mn_yTe_3)_{0.35}$ has $p = 2.09 \times 10^{20}$ carriers/cm$^3$ and $(Mn_{1+y}Sb_{2-y}Te_4)_{0.92}(Sb_{2-y}Mn_yTe_3)_{0.08}$ has $p = 9.06 \times 10^{19}$ carriers/cm$^3$.

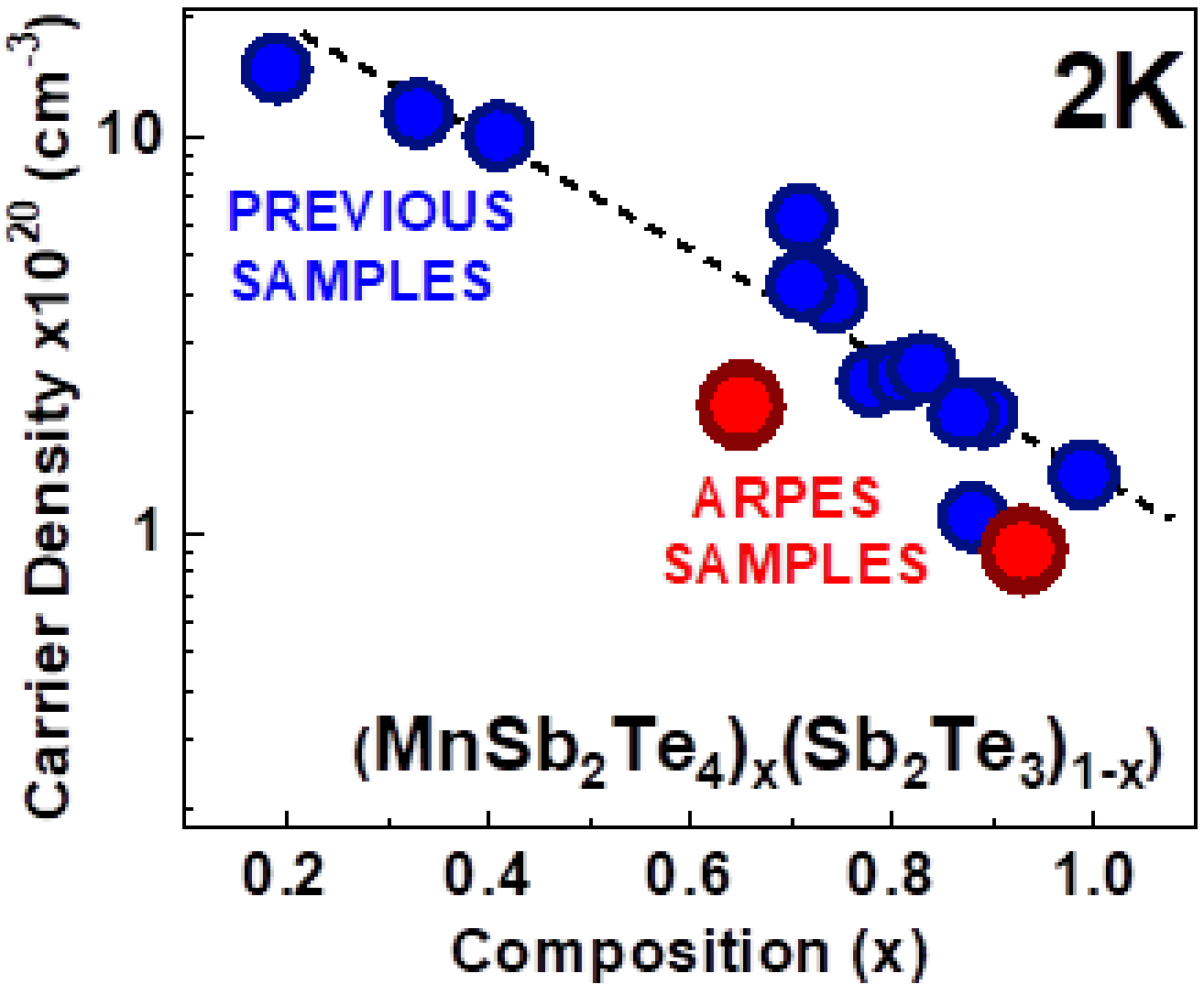

**Figure 6: a) Carrier density vs composition plot of MBE grown $(Mn_{1+y}Sb_{2-y}Te_4)_x(Sb_{2-y}Mn_yTe_3)_{1-x}$ thin films.** Blue points represent thin films previously grown with various compositions x (ref. 31). Higher x values yielded lower carrier densities. Red points are the samples used in this study.

The band structures for these two samples measured at 77eV are compared in Figure 7. Previous ARPES studies have shown that the addition of Bi in $MnSb_2Te_4$ to form $Mn(Bi,Sb)_4Te_7$ modified the carrier density of $MnSb_2Te_4$ and shifted the Fermi level 120meV closer to the charge neutrality point, revealing the apex of the Dirac cone; the observed shift showed good agreement with DFT calculations.[13] In a similar manner, increasing the composition of the $(Mn_{1+y}Sb_{2-y}Te_4)_x(Sb_{2-y}Mn_yTe_3)_{1-x}$ samples from x = 0.65 to x = 0.92 reveals a shift in the Fermi level, in our case without the addition of Bi, and allowing the onset of the Dirac point to be observed in $(Mn_{1+y}Sb_{2-y}Te_4)_{0.92}(Sb_{2-y}Mn_yTe_3)_{0.08}$, as indicated by the red dotted lines (Figures 7a and 7b). In summary, we show that a higher value of x in $(Mn_{1+y}Sb_{2-y}Te_4)_x(Sb_{2-y}Mn_yTe_3)_{1-x}$ resulted in a lower carrier density, and as expected, shifted the Fermi level higher in the valence band, making it easier to observe higher points in the valence band, and allowing to discern the onset of the Dirac cone.

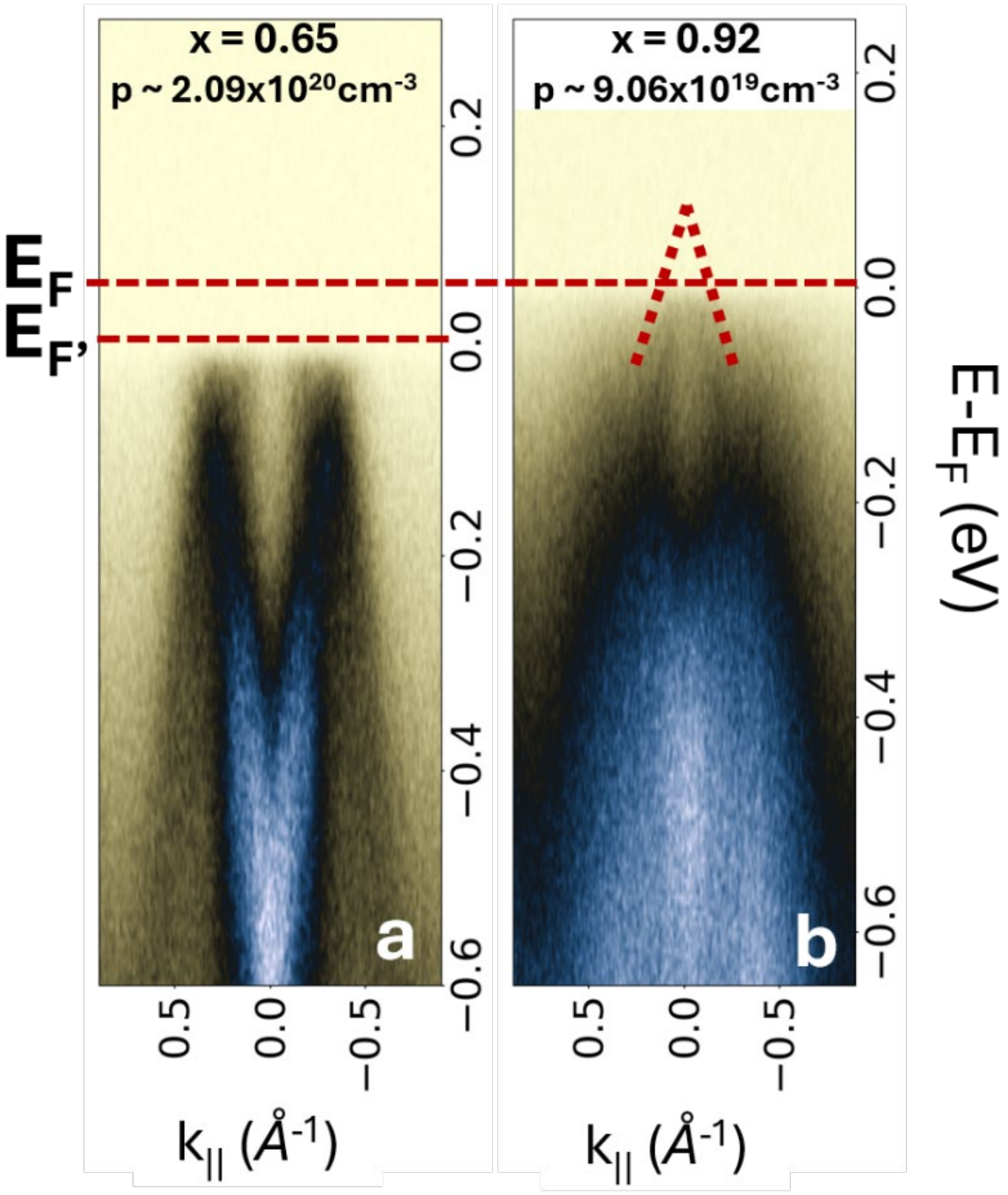

**Figure 7: ARPES study investigating effects of composition on the band structure in $MnSb_2Te_4$ based materials** **a**) Experimental ARPES band structure of $(Mn_{1+y}Sb_{2-y}Te_4)_{0.65}(Sb_{2-y}Mn_yTe_3)_{0.35}$. b) Experimental ARPES band structure of $(Mn_{1+y}Sb_{2-y}Te_4)_{0.92}(Sb_{2-y}Mn_yTe_3)_{0.08}$.

## 4 Effect of ferromagnetic interactions on the band structure

The transport properties for the two samples in our study were probed as a function of temperature. The 2K field dependent Hall resistance $R_{xy}$, shown in Figure 8a, clearly illustrates the ferromagnetic behavior of the $(Mn_{1+y}Sb_{2-y}Te_4)_{0.65}(Sb_{2-y}Mn_yTe_3)_{0.35}$ sample. The red and blue curves, measured at 60 and 70 K, respectively, show that ferromagnetic behavior persists at these high temperatures, suggesting $T_C$ values as high as 70K. Ferromagnetic behavior at temperatures as high as 100K has been observed in other samples. To investigate the effect of the magnetic order on the band structure, ARPES measurements were made as a function of temperature. Figures 9 a-f, show the band structure for high $T_C$ $(Mn_{1+y}Sb_{2-y}Te_4)_{0.65}(Sb_{2-y}Mn_yTe_3)_{0.35}$ measured at different temperatures between 18K and 120K. A reduction in the number of bands was reported by others as the temperature shifts through the $T_C$ in $MnBi_2Te_4$.[38] At first glance, no significant changes were observed in our data (Figures 9 a-f). However, upon closer analysis of the energy distribution curves some changes in the band structure as a function of binding energy could be perceived. The binding energy, measured along the dotted line of Figure 9a, shows a shift in the peak binding energy as the temperature increases, as shown in Figure 9g. This shift in binding energy ($\Delta E$ - $E_F$) is plotted as a function of temperature in Figure 9h. The binding energy data shows a downward slope, with a very large step at around 25K and a smaller but still discernible larger step at ~ 60-70K, highlighted by the green bands. These two temperature values correlate well with the expected values for the $T_{C1}$ and $T_{C2}$ for this high $T_C$ $(Mn_{1+y}Sb_{2-y}Te_4)_{0.65}(Sb_{2-y}Mn_yTe_3)_{0.35}$ sample, indicated by the black arrows in the temperature dependent $R_{xy}$ plot (blue curve), suggesting a change in the band structure as the temperature crosses the two Curie temperatures.[28,29] Changes in the band structure as the sample moves through important magnetic transitions were reported by others, too.[38,39] It is also noted that the ARPES EDC curves continue to change beyond the $T_{C2}$ value. This could be due to persistent short-range magnetic order (fluctuating local moments). Continuous band shifting past the secondary transition confirms a robust short-range spin-fluctuation regime in $MnSb_2Te_4$ (MST). Although long-range order vanishes beyond the two Curie temperatures, significant $Mn_{Sb}$ mixing may preserve local magnetic clusters active far above these temperatures.[16,40] ARPES captures these fluctuating exchange fields before they time-average out, driving the smooth energy drift in the very broad EDC signal.[41,42] More highly resolved temperature studies would be needed to fully corroborate this result.

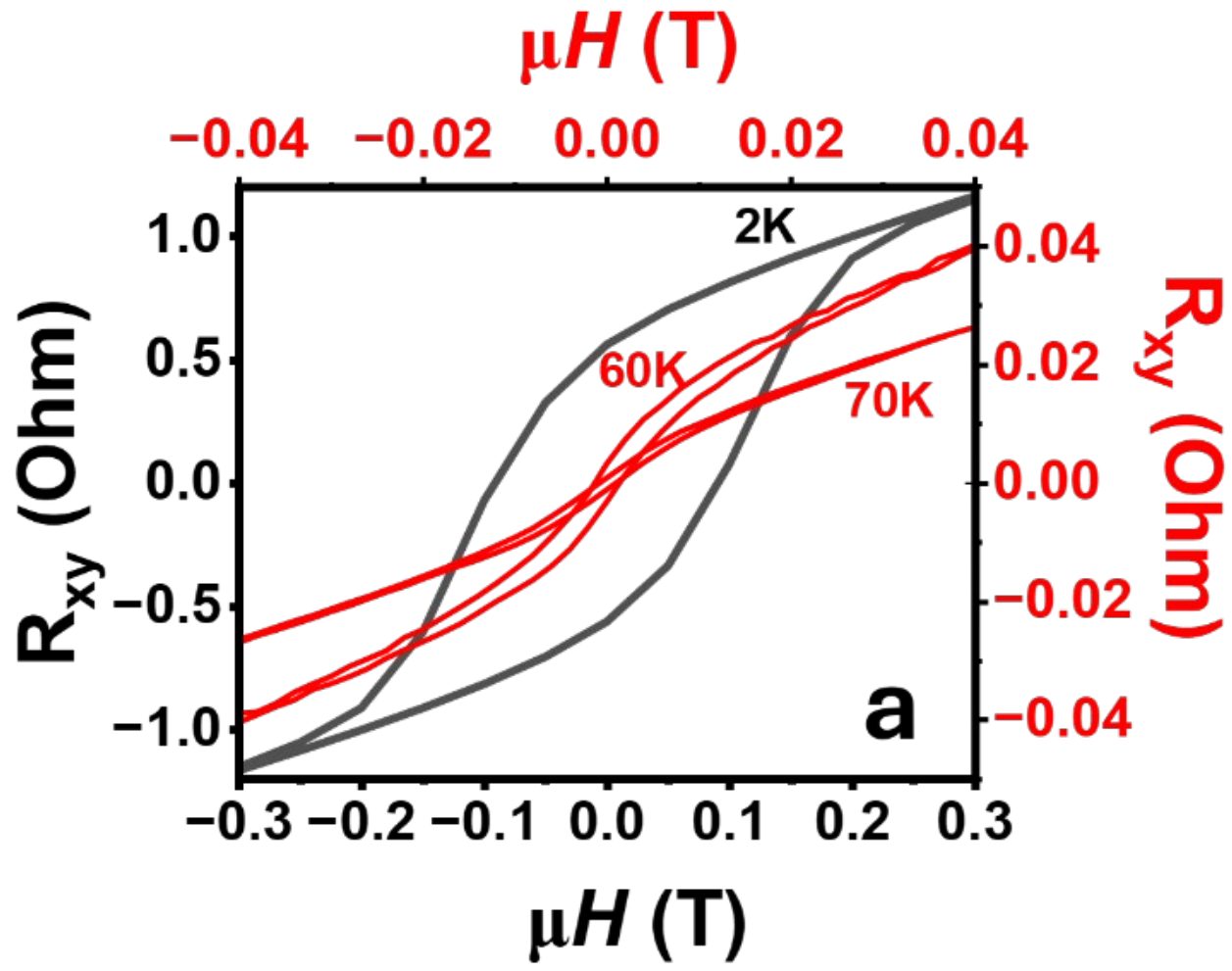

**Figure 8: Transport properties of $(Mn_{1+y}Sb_{2-y}Te_4)_{0.65}(Sb_{2-y}Mn_yTe_3)_{0.35}$.** Field dependent $R_{xy}$ measurements exhibiting a hysteresis loop, indicative of ferromagnetic behavior. Hysteretic behavior persists up to 70K.

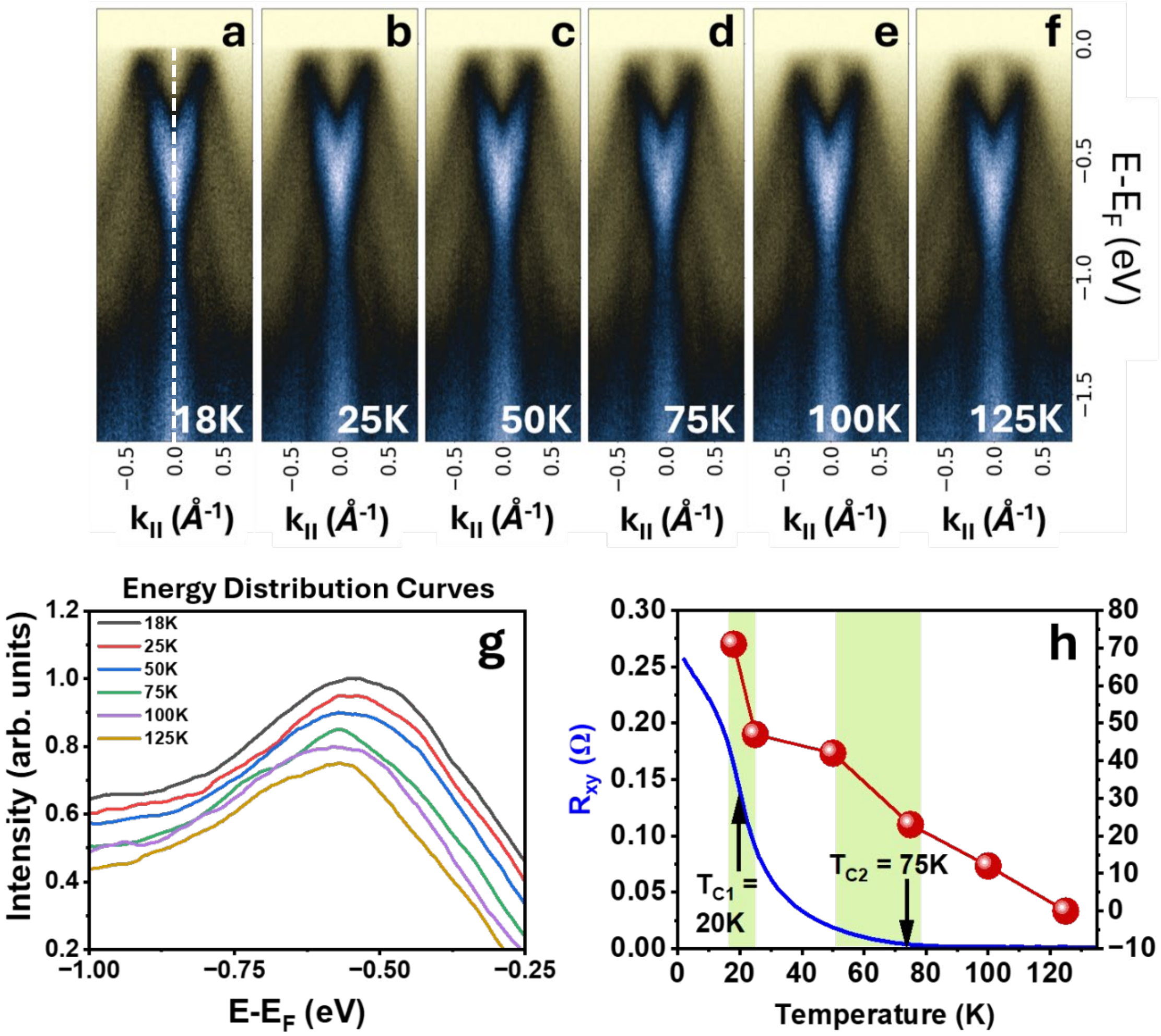

**Figure 9: Temperature dependent electronic band structure analysis of $(Mn_{1+y}Sb_{2-y}Te_4)_{0.65}(Sb_{2-y}Mn_yTe_3)_{0.35}$ sample thin film between 18K – 125K.** a-f) Images of electronic band structure taken at 77eV taken between 18K-125K. Dotted line shows where energy distribution curves were taken. g) energy distribution curves (EDC) of electronic band structure between 18K-125K where a shift towards higher binding energy is observed at lower temperatures. h) Change in binding energy ($E-E_F$) as a function of temperature (red dots). Significant step changes in $E-E_F$ (highlighted by the green bands) correspond in temperature to two possible magnetic transitions, $T_{C1}$ and $T_{C2}$, which are indicated by black arrows in the $R_{xy}$ plot (blue curve).

## 5 Conclusions

This work presents ARPES studies that explore the band structure and topological nature of two samples of high $T_C$ $(Mn_{1+y}Sb_{2-y}Te_4)_x(Sb_{2-y}Mn_yTe_3)_{1-x}$ with x = 0.65 and x = 0.92, respectively. Photon energy dependent ARPES measurements were conducted between 23eV to 83 eV. Well-defined band structure features could be seen, comparable to those reported by others for similar materials. Since the $(Mn_{1+y}Sb_{2-y}Te_4)_x(Sb_{2-y}Mn_yTe_3)_{1-x}$ materials are highly p-type doped, the band structure only probed the valence band. Absence of Fermi arcs in the Fermi surface mapping ruled out WSM behavior. Our studies revealed features that did not show dispersion as a function of the photon energy, consistent with the behavior of surface bands, supporting the non-trivial nature of the materials. Further evidence of their topological surface nature was provided by $k_x$ vs $k_z$ studies.

Our results also showed that by varying the composition (i.e., the value of x) of the materials we were able to tune the Fermi level. moving the Fermi level slightly higher in the valence band in the sample with x = 0.92. As the Fermi level shifted toward the gap, the onset of the Dirac cone could be seen, consistent with a topological insulator behavior. Temperature dependent ARPES studies revealed possible changes in the band structure as the temperature crossed the magnetic transition temperatures of our materials. More systematic studies of the temperature dependence may allow us to better understand the nature of the magnetic interactions. Some mechanisms that have been proposed to describe how bands or peaks merge as a function of temperature include the Stoner mechanism and the localized exchange mechanism.[43]

This work also demonstrates that that thin film samples prepared using a Te cap, which is easily desorbed prior to the ARPES measurement, can be successfully applied to yield valuable band structure information, and shows that ARPES is a very useful technique to investigate the band structure of these highly disordered and complex topological materials.

## 6 Acknowledgements

This work was supported by NSF Grant no. DMR-2011738 (NSF MRSEC PAQM) and NSF Grant no. HRD-2112550 (Phase II CREST IDEALS). The authors would like to acknowledge the Nanofabrication Facility of the CUNY Advanced Science Research Center (ASRC) for instrument use and scientific and technical assistance and the staff at Brookhaven National Laboratory (BNL). This research used the ESM beamline of the National Synchrotron Light Source II, a U.S. Department of Energy (DOE) Office of Science User Facility operated for the DOE Office of Science by Brookhaven National Laboratory under Contract No. DE-SC0012704.

The authors acknowledge the helpful conversations with Dr. Kyungwha Park that provided interpretations of the ARPES characterization.

# Elucidating the topological nature of molecular beam epitaxy grown high Curie temperature $(Mn_{1+y}Sb_{2-y}Te_4)_x(Sb_{2-y}Mn_yTe_3)_{1-x}$ using angle resolved photoelectron spectroscopy

Candice R. Forrester[1,2], Jisoo Moon[1], Sina Mohammadi[3,4], Asish Kundu[5], Amy Lopez Navarro[1], Kaushini Wickramasinghe[6], Abhay Pasupathy[5,7], Anil K. Rajapitamahuni[5], Maya N. Nair[5,8], Elio Vescovo[5], Lia Krusin-Elbaum[4,9], Maria C. Tamargo[1,2,4,9, a]

*[1]Department of Chemistry, The City College of New York, NY, NY 10031, USA*

*[2]Ph.D Program in Chemistry, CUNY Graduate Center, NY, NY 10016, USA*

*[3]Department of Physics, Queens College of CUNY, NY, NY 10031, USA*

*[4]Ph.D Program in Physics, CUNY Graduate Center, NY, NY 10016, USA*

*[5]Brookhaven National Laboratory, Upton, NY 11973, USA*

*[6] Department of Physics, University of Peradeniya, Peradeniya, 20400, Sri Lanka*

*[7]Department of Physics, Columbia University, New York, NY 10027, USA*

*[8]CUNY Advanced Research Science Center, New York, NY 10031, USA*

*[9]Ph.D Program in Physics, CUNY Graduate Center, NY, NY 10016, USA*

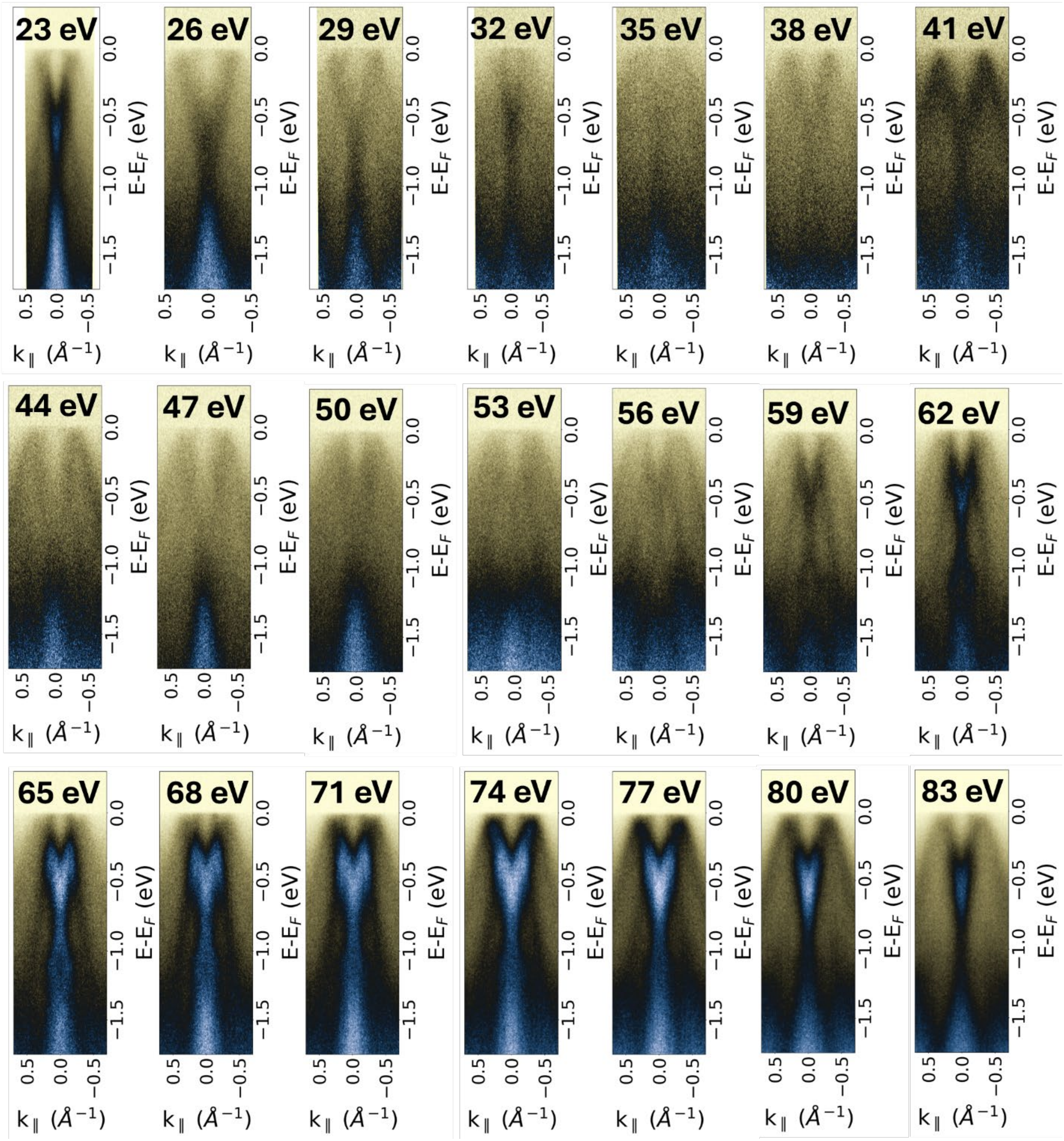

**Figure S1: Photon energy dependent study on $(MnSb_2Te_4)_{0.65}(Sb_2Te_3)_{0.35}$**

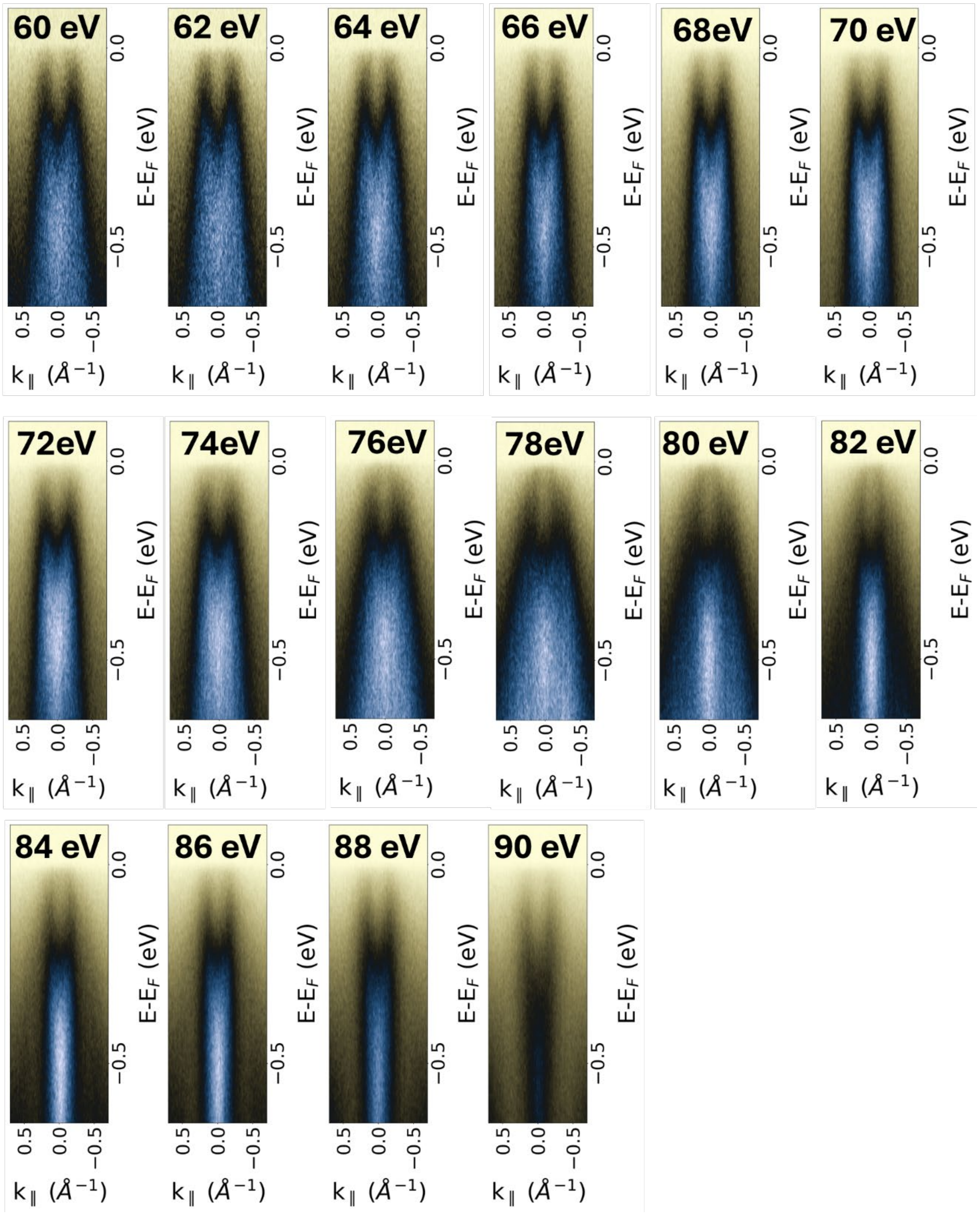

**Figure S2: Photon energy dependent study on $(MnSb_2Te_4)_{0.92}(Sb_2Te_3)_{0.08}$**